\documentclass{article}

\PassOptionsToPackage{numbers, compress}{natbib}
\usepackage{arxiv}

\usepackage[utf8]{inputenc} 
\usepackage[T1]{fontenc}    
\usepackage{hyperref}       
\usepackage{url}            
\usepackage{booktabs}       
\usepackage{amsfonts}       
\usepackage{nicefrac}       
\usepackage{microtype}      

\usepackage{amsmath}
\usepackage{amssymb}
\usepackage{xfrac}          
\usepackage{comment}        
\usepackage{multicol}       
\usepackage{multirow}
\usepackage[inline]{enumitem}  
\usepackage{tabularx}       
\usepackage[ruled]{algorithm2e}
\usepackage{caption}

\DeclareMathOperator{\logSumExp}{LSE}

\newcommand{\q}{q}
\newcommand{\p}{p}
\newcommand{\GS}{\text{RelaxedCategorical}}
\newcommand{\DP}{p}  
\newcommand{\KLDGS}{D_{KL}}
\newcommand{\EE}{\mathbb{E}}
\newcommand{\sumN}[1]{\sum_{{#1}=1}^n}
\newcommand{\lse}[2]{\logSumExp_{{#1}=1}^n \left( #2 \right)}

\newcommand{\z}[1]{\frac{\log \alpha_{#1} + g_{#1}}{\lambda}}
\newcommand{\llamb}{\frac{l}{\lambda}}

\title{ReCAB-VAE: Gumbel-Softmax\\ Variational Inference based on Analytic Divergence}

\author{%
  Sangshin Oh \\
  Yonsei University\\
  \texttt{sangshin.oh@dsp.yonsei.ac.kr} \\
  \And
  Seyun Um \\
  Yonsei University \\
  \texttt{syum@dsp.yonsei.ac.kr} \\
  \And
  Hong-Goo Kang \\
  Yonsei University  \\
  \texttt{hgkang@yonsei.ac.kr} \\
}

\begin{document}
    
\maketitle

\begin{abstract}
The Gumbel-softmax distribution, or Concrete distribution, is often used to relax the discrete characteristics of a categorical distribution and enable back-propagation through differentiable reparameterization. Although it reliably yields low variance gradients, it still relies on a stochastic sampling process for optimization. In this work, we present a relaxed categorical analytic bound (ReCAB), a novel divergence-like metric which corresponds to the upper bound of the Kullback-Leibler divergence (KLD) of a relaxed categorical distribution. The proposed metric is easy to implement because it has a closed form solution, and empirical results show that it is close to the actual KLD. Along with this new metric, we propose a relaxed categorical analytic bound variational autoencoder (ReCAB-VAE) that successfully models both continuous and relaxed discrete latent representations. We implement an emotional text-to-speech synthesis system based on the proposed framework, and show that the proposed system flexibly and stably controls emotion expressions with better speech quality compared to baselines that use stochastic estimation or categorical distribution approximation.
\end{abstract} 

\section{Introduction}
    Inspired by recent successes in generating high fidelity images or audio using deep neural networks, modeling and manipulating specific class information and its fine attributes has become an active research area \citep{mirza2014conditional, kingma2014semi, sohn2015learning, van2016conditional, luong2017adapting, choi2018stargan}. Since many attributes of real world data such as object identity, gender, and emotion are often annotated categorically, a natural choice is to model them using categorical or discrete latent representations. When the discrete structure of a data set is modeled in this manner, the latent representation can be easily interpreted since each dimension represents one of the classes or categories. For example, conditional generative adversarial networks (cGANs) \citep{mirza2014conditional} and conditional variational autoencoders (cVAEs) \citep{sohn2015learning} take categorical auxiliary inputs, or \emph{conditions}, which are fed to both the discriminator (or recognition) network and generator network to control the attributes of the generated data.
    
    An alternative approach was proposed by \citet{kingma2014semi} in which a generative model was expanded to a variational approach in a semi-supervised training framework. 
    Here, it is possible to infer the class label by using an auxiliary encoder even when the annotated label is not available. The auxiliary encoder estimates the probability that a sample belongs to each category, after which the estimated probability is used to calculate the marginal likelihood of the sample.
    Recently, the Gumbel-softmax distribution was proposed as a differentiable relaxation for discrete or categorical variables \citep{jang2016categorical, maddison2016concrete}. In this framework, unannotated labels are directly inferred using relaxed categorical variables and back-propagated to the encoder using a reparameterization trick. This allows for a significant reduction in inference time because it does not need a marginalization over all possible latent variables, unlike the semi-supervised method.

    The relaxed categorical distribution is frequently applied for variational inference \citep{jang2016categorical, maddison2016concrete, dupont2018learning, yang2017improved, siddharth2017learning, kipf2018neural}, in which a distribution of latent representation estimated from observed data (a variational posterior) needs to be fitted to the prior distribution. The Kullback-Leibler divergence (KLD) \citep{kullback1951information} plays an important role in this optimization process because it quantitatively measures the similarity between two distributions. The calculation of the KLD is easy in the Gaussian distribution case because of the existence of a simple analytic solution. However, no analytic solution exists for the relaxed categorical distribution case; therefore, the KLD must be calculated from a stochastic estimation process or approximated using a normal categorical distribution.
    
    In this work, we solve this problem by calculating the upper bound of the KLD using a closed-form solution, which is more accurate than conventional stochastic estimation or distribution approximation methods. Specifically, we propose a relaxed categorical analytic bound variational autoencoder (ReCAB-VAE), which has two distinctive features from previously suggested VAEs or the jointVAE framework \citep{jang2016categorical, maddison2016concrete, dupont2018learning}. The relaxed categorical analytic bound (ReCAB), which is the training objective for ReCAB-VAE, has a very similar value to the true KLD of relaxed categorical distributions, but can be calculated analytically and efficiently. The proposed metric is lower-bounded by the true KLD, so it can be easily used to optimize the variational bound. In other words, unlike conventional approximation-based approaches, the proposed metric reliably sets a lower bound for the logarithmic evidence in the training objective. In addition, we also propose a \emph{discriminative prior} that utilizes the information from a class label of the data. Unlike uninformative priors such as the standard Gaussian or uniform priors, the proposed discriminative prior considers a class label to determine its corresponding parameters. By giving distinctive information to its prior, it prevents the variational posterior from collapsing into a non-informative uniform distribution. This means that we do not need to provide any additional classification loss when training the model because the latent representations are distinguishable due to the guidance of the discriminative prior.

    To show that our model is able to generate high fidelity samples while flexibly controlling their attributes, we implement an emotional text-to-speech (TTS) synthesis system with the proposed framework. An emotional TTS system is a suitable application for this framework because it needs to faithfully control emotional expressions as well as textual content. The textual content is represented in the continuous space and the emotional expressions are often annotated with class labels in the discrete latent space \citep{lee2017emotional, habib2019semi, kwon2019effective}. Experimental results show that the synthesized speech quality of the proposed system is much higher than that of conventional ones that use stochastic estimation or categorical distribution approximation. In addition, the proposed system stably controls emotional content while faithfully maintaining the textual content.
    
    The rest of this paper is organized as follows. In Section \ref{sec:pre}, we briefly describe the fundamentals of the relaxed categorical distribution and the jointVAE algorithm that uses both continuous and discrete latent spaces. In Section \ref{sec:body}, we explain the proposed ReCAB-VAE model in detail. Experiments and results are described in Section \ref{sec:exs}, and conclusions follow in Section \ref{sec:conc}.

\section{Preliminaries}
\label{sec:pre}
    We first describe the relaxation of categorical or discrete variables which enables direct back-propagation through the stochastic node \citep{jang2016categorical, maddison2016concrete}. Then, we review the jointVAE algorithm that was recently developed to improve the performance of conventional VAE approaches \citep{dupont2018learning}.
    
    \subsection{Relaxation of categorical random variables}
    Motivated by the Gumbel-Max trick \citep{gumbel1954statistical}, \citet{jang2016categorical} and \citet{maddison2016concrete} concurrently suggested a relaxation method for categorical random variables. If a categorical variable takes the class probability of $(\alpha_1, \alpha_2, \dots, \alpha_n)$ for each category, a relaxation of this variable is defined as the softmax of $\sfrac{\log \alpha_k + g_k}{\lambda}$, where $g_k$ is independent standard Gumbel noise and $\lambda$ is the temperature parameter. Precisely, the $k$-th element of latent variable $z$ is defined as:
    \begin{align}
        z_k = \frac{\exp \left( \sfrac{(\log \alpha_k + g_k)}{\lambda} \right)}{\sumN{i} \exp \left( \sfrac{(\log \alpha_i + g_i)}{\lambda} \right)} \label{eq:RV}.
    \end{align}
    Since the latent variable is properly reparameterized with the Gumbel noise, the gradient can be calculated and back-propagated through this stochastic node. This relaxation process is considered to be a generalized version of the original categorical distribution because it converges to the exact categorical distribution when the temperature $\lambda$ goes to 0.
    The probability density function of this relaxation process is defined as follows:
    \begin{align}
        p (z) = \Gamma(n) \lambda^{n-1}\prod_{k=1}^n \frac{\alpha_k z_k^{-\lambda -1}}{\sumN{i} \alpha_i z_i^{-\lambda}} \label{eq:density},
    \end{align}
    where $\Gamma(\cdot)$ is a Gamma function.
    In this paper, we use $\alpha$ and $\lambda$ to denote the logit and temperature parameters of the relaxed categorical distribution, respectively.
    
    \subsection{JointVAE framework}
    Recently, \citet{dupont2018learning} proposed the jointVAE model, which embeds its inputs in both continuous and discrete latent spaces. Here, it was shown that it is beneficial to model some attributes in a disentangled space from the continuous latent space. The continuous and discrete latent variables are assumed to be independent of each other, so the objective of this variational model is given as follows:
    \begin{align}
        \EE_{q_\phi (z_c, z_d | x)} \big[ \log p_\theta (x | z_c, z_d) \big]
            - \KLDGS \left( q_\phi (z_c|x) \| p(z_c) \right) - \KLDGS \left( q_\phi (z_d | x) \| p(z_d) \right),
    \end{align}
    where $z_c$ and $z_d$ refer to the continuous and discrete latent variables, respectively, and $\phi$ and $\theta$ denotes encoder and decoder parameters.

\section{Relaxed Categorical Analytic Bound Variational Autoencoder}
\label{sec:body}
    \subsection{Relaxed Categorical Analytic Bound}
    We propose a relaxed categorical analytic bound (ReCAB), which is lower-bounded by the KLD of the relaxed categorical distribution. To this end, we decompose the equation from the definition of the KLD into a summation of several expectation terms. 
    Then, we reformulate or bound each term so that the bound can be solved in a closed form. We first provide a brief mathematical derivation for our proposed bound, and then compare it with conventional estimation or approximation approaches for the KLD.
        
    \subsubsection{Derivation of the bound}
    From the density function in Eq. (\ref{eq:density}), the KLD of the relaxed categorical distributions from prior $\p (z) = \GS(a, l)$ to posterior $\q (z|x) = \GS(\alpha, \lambda)$ is defined as follows:
    \begin{align} \begin{split}
        \KLDGS \left( \q (z|x) \| \p (z) \right)
            &= \EE_{z} \big[ \log \q (z|x) - \log \p (z) \big] \\
            &= \EE_{z} \big[ (n-1) \log \frac{\lambda}{l} + \sumN{k} \bigg( (\log \alpha_k - \lambda \log z_k) - (\log a_k - \lambda \log z_k) \bigg) \\
            & \qquad  {} - n \lse{i}{\log \alpha_i - \lambda \log z_i} + n \lse{i}{\log a_i - \lambda \log z_i} \big],
    \end{split} \end{align}
    where $z_k$ is the $k$-th element of discrete latent variable $z_d$ and $\logSumExp$ denotes a log-sum-exponential operator.
    In here, $a$ and $l$ are the logit and temperature parameters of the prior distribution, while $\alpha$ and $\lambda$ denote those of the posterior.
    Note that terms that appear in both the posterior and prior cancel out.
    For notational simplicity and clarity, we omit the subscript of the discrete latent variable $z$ and succinctly write $\EE_{z \sim \q (z|x)}$ as $\EE_{z}$. 
    In order to simplify the above equation, we can substitute logit terms with the following equation, which is derived from Eq. (\ref{eq:RV}).
    \begin{align} \begin{split}
        \log \alpha_k - \lambda \log z_k
            &= - g_k + \lambda \lse{i}{\sfrac{(log \alpha_i + g_i)}{\lambda}} \\
        \log a_k - l \log z_k
            &= \log a_k - \frac{l}{\lambda} \log \alpha_k + l \lse{i}{\sfrac{(log \alpha_i + g_i)}{\lambda}}.
    \end{split} \end{align}
    This results in the following equation:
    \begin{align} \begin{split}
        \KLDGS \left( \q (z|x) \| \p (z) \right)
            &= - (n-1) \log \frac{l}{\lambda} - \sumN{k} A_k - (1-\frac{1}{\lambda}) \EE_g \big[ \sumN{k} g_k \big] \\
            & \qquad {} - n \cdot \EE_g \big[ \lse{k}{g_k} \big] + n \cdot \EE_g \big[ \lse{k}{A_k - \frac{l}{\lambda} g_k} \big],
    \end{split} \end{align}
    where $A_k = (\log a_k - \frac{l}{\lambda} \log \alpha_k)$.
    Since the last term cannot be solved analytically, we apply Jensen's inequality:
    \begin{align} \begin{split}
        \EE_g \big[ \lse{k}{A_k - \frac{l}{\lambda} g_k} \big]
            &\le \log \EE_g \big[ \sumN{k} \left( \exp A_k \cdot \exp ( - \frac{l}{\lambda} g_k ) \right) \big] \\
            &= \log \sumN{k} \left( \exp A_k \cdot \EE_g \big[ \exp ( - \frac{l}{\lambda} g_k ) \big] \right),
    \end{split} \end{align}
    where the last RHS term is obtained by reordering the expectation and summation operations.
    Since the expectation term becomes a constant value once evaluated, the upper bound of the last RHS term can be solved analytically.
    By calculating all of the remaining expectations, it is possible to obtain the upper bound of the KLD as follows:
    \begin{align} \begin{split} \label{eq:finalKLD}
        D_{ReCAB} \big( \q(z|x) \| p(z) \big)
            &= - (n-1) \log \frac{l}{\lambda} + n \left( \gamma \frac{l}{\lambda} + \log \Gamma (1 + \frac{l}{\lambda}) - \sum^{n-1}_{k=1} \frac{1}{k} \right) \\
            & \qquad {} - \sumN{k} \text{log-softmax}_k ( \log {a} - \frac{l}{\lambda} \log {\alpha} ) \\
            &\ge \KLDGS \left( \q (z|x) \| \p (z) \right).
    \end{split} \end{align}
    Since the last $\text{log-softmax}(\cdot)$ term is the only term that is related to logits $\alpha$ and $a$, the other terms in Eq. (\ref{eq:finalKLD}) can be calculated in advance when the temperatures are fixed throughout training. 
    Therefore, it is easy to implement a training algorithm because we only need to evaluate one function at each epoch. The general algorithm for training is depicted in Algorithm \ref{algo:train}. 
        
    Detailed derivations are provided in the supplementary material.
    
    \subsubsection{Comparison with conventional approaches}
    In previous works, the KLD of the relaxed categorical distribution was estimated via the Monte-Carlo method (Eq. (\ref{eq:kldMCE}), MC) or roughly approximated by using the KLD of the categorical distribution (Eq. (\ref{eq:kldRA}), CA) as follows:
    \begin{align}
        \KLDGS \left( \q (z|x) \| \p (z) \right)
            &\approx \sum^{L}_{i=1} \bigg( \log \q (z^{(i)} | x) - \log \p (z) \bigg) = D_{MC} \label{eq:kldMCE} \\
            &\approx \sumN{k} \hat{\alpha}_k \bigg( \log \hat{\alpha}_k - \log {a}_k \bigg) = D_{CA} \label{eq:kldRA},
    \end{align}
    where $L$ is the number of samples for stochastic estimation and $\hat{\alpha}_k$ denotes the $k$-th element of the normalized logit parameter or categorical probability, $\hat{\alpha}_k = \sfrac{\alpha_k}{\sumN{i} \alpha_i}$. 
    Eq. (\ref{eq:kldMCE}) must be evaluated from the density function given in Eq. (\ref{eq:density}), so it is somewhat complicated to calculate. In addition, it provides noisy and inaccurate values for the KLD when $L$ is not large enough. 
    Because of this complexity and inaccuracy, it is a tempting choice to use the categorical distribution to approximate the KLD as in Eq. (\ref{eq:kldRA}). Unlike stochastic estimation, this gives the approximated value in a deterministic and efficient way.
    However, because the categorical distribution does not have any temperature or scale parameters, this approximation cannot consider the temperature parameter of the relaxed categorical distribution. Therefore, this method only provides a rough approximation for the KLD which is inaccurate and results in an irregular loss landscape. In addition, as pointed out in previous works \citep{jang2016categorical, maddison2016concrete}, the training objective with this approximation is not guaranteed to be an evidence lower bound.
    
    \subsection{Discriminative prior}
    In this section, we propose a \emph{discriminative prior} which can be used for semi-supervised or supervised training of ReCAB-VAE. Unlike the uniform distribution or its relaxed variants (e.g., a distribution with $\alpha = [\frac{1}{n}, \frac{1}{n}, \cdots, \frac{1}{n}]$, $\lambda \in (0, \infty)$), this prior provides some information about data points as in \citet{sohn2015learning} or \citet{casale2018gaussian}. 
    Since the prior distribution is now conditioned on auxiliary code or label $c$, the evidence lower bound is revised as follows:
    \begin{align} \begin{split} \label{eq:discreteElbo}
        \mathcal{L}_{ELBo} &= \EE_{z} \big[ \log p_{\theta}(x, z | c) - \log q_{\phi}(z|x) \big] \\
            &= \EE_{z} \big[ \log p_{\theta} (x | z) \big] - \KLDGS \big( q_{\phi}(z|x) \| \DP(z | c) \big),
    \end{split} \end{align}
    where $\DP(z|c)$ is the proposed discriminative prior. Note that we assume the code $c$ to be independent of both the encoder $q_\phi$ and decoder $p_\theta$ here. The proposed prior is a usual relaxed categorical distribution whose logit $a$ is a function of the auxiliary code $f_a(c;\epsilon)$. The function for the logit parameter can be defined as follows:
    \begin{align} \begin{split}
        f_a (c;\epsilon) = 
        \begin{cases}
            1 - (n-1)\epsilon & \text{if } c_i = 1 \\
            \epsilon          & \text{otherwise}
        \end{cases},
    \end{split} \end{align}
    where $c \in \{0,1\}^n$ is a one-hot discrete code and $\epsilon > 0$ is a small smoothing constant. The resulting prior is defined as $p(z|c) = \text{GumbelSoftmax} (f_a(c; \epsilon), l)$,
    where the temperature $l$ is selected freely and usually fixed throughout training.
    Using this prior, the model can learn distinctive latent representations for each class without auxiliary classification loss, unlike in \citet{kingma2014semi} or \citet{habib2019semi}.

    \subsection{Model structure}
    \begin{algorithm}[t]
        \SetAlgoLined
        \caption{Training procedure for ReCAB-VAE. Note that Encoder is separated into continuous and discrete parts in a real implementation. $\mathcal{L}_1$ can be ignored in parameter updates since it is a constant throughout the training procedure.
        }
        $\theta$, $\phi$ $\gets$ Initialize parameters \\
        $l$, $\lambda$ $\gets$ Set temperatures for prior and posterior distributions \\
        $\mathcal{L}_1 $ $\gets$ $-(n-1) \log \frac{l}{\lambda} + n\gamma \frac{l}{\lambda} 
            + \log \Gamma (1+\frac{l}{\lambda}) - \sum^{n-1}_{k=1}\frac{1}{k}$ \\
        \While {not converges}{
            $\mu$, $\sigma$, $\alpha$ $\gets$ Encoder($x$; $\phi$) \\
            $z_c$, $z_d$ $\gets$ Sample via reparameterization trick \\
            $\hat{x}$ $\gets$ Decoder($z_c$, $W_d z_d$; $\theta$) \\
            $\mathcal{L}_{2}$ $\gets$ $
                \EE_{} \big[ \log p_\theta (x | z_c, z_d) \big]
                - D_{KL} \big( q_\phi (z_c | x) \| \DP (z_c) \big)
                -\text{log-softmax} (\log a - \frac{l}{\lambda} \log \alpha$) \\
            $\mathcal{L}$ $\gets$ $\mathcal{L}_1 + \mathcal{L}_2$ \\
            $\theta$, $\phi$ $\gets$ Update parameters with $\nabla_{\theta, \phi} \mathcal{L}_2$
        }
    \label{algo:train}
    \end{algorithm}

    In this section, we propose ReCAB-VAE, a relaxed categorical analytic bound variational autoencoder based on ReCAB and the aforementioned discriminative prior. 
    Although we apply our proposed features in a VAE framework, the features explained above can be adopted in other frameworks that use the relaxed categorical distribution. Notably, ReCAB can be used as a closed-form surrogate for the KLD of the relaxed categorical distribution.
    
    ReCAB-VAE consists of two independent latent encoders as in jointVAE \citep{dupont2018learning}: a continuous latent and a discrete latent encoder. Assuming independence between the continuous and discrete latent variables, the training objective $\mathcal{L}_{Prop}$ is written as follows:
    \begin{align} \begin{split}
        \mathcal{L}_{Prop}
            &= \EE_{} \big[ \log p_\theta (x | z_c, z_d) \big] 
                - D_{KL} \big( q_\phi (z_c|x) \| \DP (z_c) \big)
                - D_{ReCAB} \big( q_\phi (z_d|x) \| \DP (z_d|c) \big) \\
            &\le \EE_{} \big[ \log p_\theta (x | z_c, z_d) \big]
                - D_{KL} \big( q_\phi (z_c | x) \| \DP (z_c) \big)
                - D_{KL} \big( q_\phi (z_d | x) \| \DP (z_d | c) \big) \\
            &= \mathcal{L}_{ELBo}
    \end{split} \end{align}
    where $c$ is an additional condition code for the discriminative prior. Note that our proposed objective is bounded above by ELBo. 
    With the learning objective specified above, a structure for the proposed model is as follows:
    \begin{align}
        [\mu; \log \sigma^2] &= \text{Cont.Encoder} (x) &
        \alpha &= \text{Disc.Encoder} (x) \nonumber \\
        z_c &\sim \mathcal{N}(\mu, \sigma^2) &
        z_d &\sim \GS (\alpha, \lambda) \\
        \hat{x} &= \text{Decoder} (z_c, W_d z_d) \nonumber,
    \end{align}
    where $\hat{x}$ is the reconstruction of sample $x$ and $\lambda$ is fixed throughout the training process as in previous works \citep{jang2016categorical, maddison2016concrete}. 
    Both latent encoders, $\text{Cont.Encoder}$ and $\text{Disc.Encoder}$, take the data sample $x$ and represent it in parallel in the latent spaces. 
    The latent variables are then sampled via the reparameterization trick.
    In order to match the dimension of continuous and discrete latent representations, we project the discrete latent variable with embedding matrix $W_d$. Then, the decoder reconstructs the data sample from the continuous latent representation $z_c$ and projected discrete latent representation $W_d z_d$.

\section{Experiments and Results}
\label{sec:exs}
    In this section, we show that ReCAB provides a better estimation of the true KLD compared to conventional stochastic and distribution approximation approaches. We also conduct experiments to evaluate the performance of an emotional text-to-speech (TTS) system that is implemented using the proposed ReCAB-VAE framework.
    
    \subsection{Basic empirical studies}
    
    \begin{figure}[t]
        \centering
        \includegraphics[width=\textwidth]{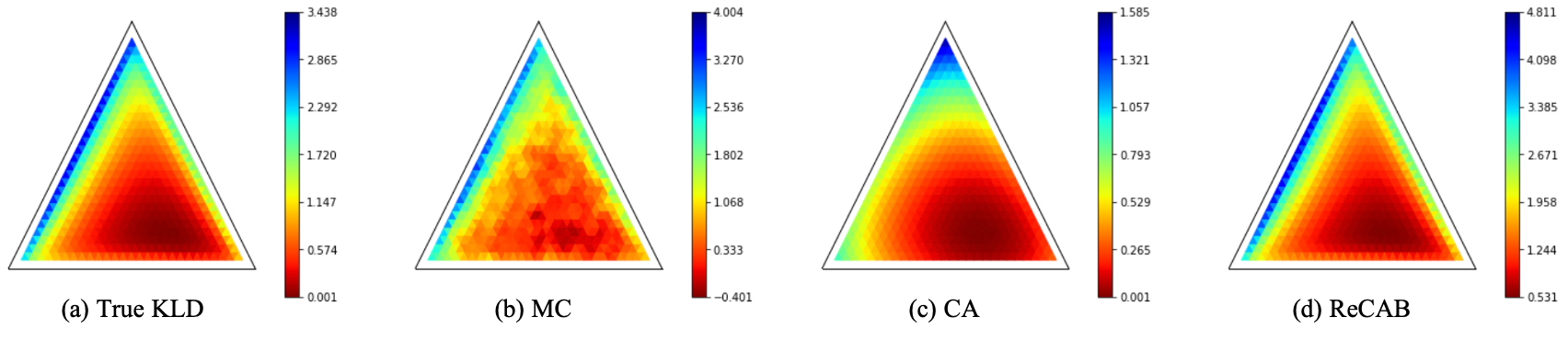}
        \caption{Heatmaps of KLD estimations from each method when the target logit is [0.57, 0.14, 0.28]. The temperatures for the target and proposal distributions are equivalently set to 1. Each point in the heatmap indicates the logit of the proposal distribution after normalization. (a) True KLD obtained from MC estimation with 300k samples, (b) MC estimation with a small batch size (32 samples), (c) CA estimation, (d) ReCAB.}
        \label{fig:heat-div}
    \end{figure}
    
    \begin{figure}[t]
        \centering
        \includegraphics[width=\textwidth]{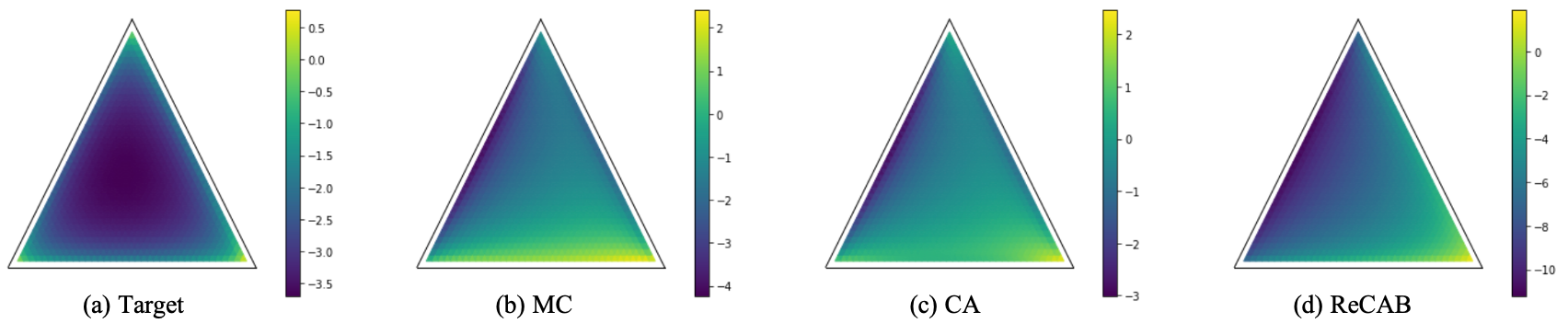}
        \caption{Density plots of target and optimum points from each method. The temperature of the target distribution was set to 0.2, while those of all other distributions were set to 1.
        (a) Target distribution with logit of [0.57, 0.14, 0.28]. Densities for optimum logit from (b) MC with 32 samples, (c) CA, and (d) ReCAB.}
        \label{fig:heat-density}
    \end{figure}
    
    Figure \ref{fig:heat-div} shows 2-dimensional simplex heatmaps that depict the true KLD (a) and estimated values from three different methods (b-d): a Monte-Carlo (MC) estimator, a categorical distribution approximation (CA) estimator, and the proposed ReCAB estimator. The three vertices of each heatmap denote the logits for the relaxed categorical distribution representing the three categories. The color at each point represents the estimated KLD with the target distribution. As shown in the figure, the estimated results obtained by the ReCAB estimator are almost identical to those computed using the exact KLD. On the other hand, MC estimation with a small batch size results in noisy outputs, and CA estimation results in a differently shaped, distorted heatmap. 
    
    Figure \ref{fig:heat-density} illustrates the density plots of optimal logit parameters from the MC, CA, and ReCAB estimators. 
    When the temperatures are set to be identical to the prior (or target), the densities from all the estimators fit well to the target distribution (shown in supplementary material).
    When the temperatures are different from the prior, however, each density plot shows distinctive results. 
    As shown in Eq. (\ref{eq:kldRA}), the categorical approximation causes the local optimum to ignore the temperature parameter; this results in a density that is very different from the target distribution.
    Meanwhile, MC estimation and ReCAB show relatively similar densities to the target distribution, but MC estimation results in a slightly worse optimum because of its noisiness.

    \subsection{Emotional speech synthesis}
    We performed experiments on emotional speech synthesis using an internal emotional TTS database consisting of 26.4 hours of speech. The database contains 21,678 utterances, where each sentence is recorded by a male and a female speaker. About one third of the utterances (6,604) are annotated with one of 4 emotional classes (\textit{angry, happy, neutral, sad}). The others are not annotated because of ambiguity in labeling. 
    
    \paragraph{Architectures.}
    To generate emotional speech with our proposed model, we chose Tacotron 2 \citep{shen2018natural} as our baseline system, an end-to-end TTS system based on an encoder-decoder framework. Similar to previous approaches for Tacotron-based controllable speech synthesis \citep{habib2019semi, skerry2018towards, wang2018style, zhang2019learning, sun2020fully}, our ReCAB-VAE model is an augmentation to the baseline that allows control over the emotional expression of the generated speech signal. In particular, we borrowed the \emph{reference encoder} architecture for the latent encoder of ReCAB-VAE. Since we have two separate encoders for continuous and discrete latent representations, we halve the number of channels in the convolutional layers and the number of cells in the recurrent layer. 
    The representations from each latent encoder are then concatenated and added to the embedding from the text encoder of Tacotron 2. 
    We compared versions of this model trained with Monte-Carlo estimation (MC), categorical distribution approximation (CA) and ReCAB.
    We used WaveGlow \citep{prenger2019waveglow} as a neural vocoder to transform the predicted mel-spectrogram into waveforms for all models. 
    
    \subsubsection{Subjective tests}
    We conducted two subjective tests to evaluate the performance of our proposed system.
    In the first subjective test, raters were asked to assess the overall quality of the synthesized speech signal on a scale from 1 to 5 to obtain the mean opinion score (MOS). The second test was an ABX test in which raters were asked to choose one of two speech samples from different models according to their preferences. Emotion labels were shown to the raters in the ABX test, while they were not shown in the MOS test.
    
    \paragraph{Mean opinion score test.}
    \begin{table}[t]
        \caption{Speech quality and naturalness mean opinion scores (MOS) with 95\% confidence intervals. \textbf{Neutral voices} indicates scores that only consider speech samples with the \textit{neutral} emotion, while all of the speech samples are considered in \textbf{Total}.
        VAEs with Monte-Carlo estimation (MC) and normal categorical distribution approximation (CA) have same architecture as ReCAB-VAE, but are trained with different training objectives.}
        \vspace{4pt}
        \label{tab:mos}
        \centering
        \begin{tabular}{l c c}
            \toprule       
              \bf{Model}   &   \bf{Neutral voices} &          \bf{Total}  \\
            \midrule       
            Recorded       &       4.84 $\pm$ 0.10 &     4.87 $\pm$ 0.05  \\
            GT-Mel         &       4.69 $\pm$ 0.08 &     4.57 $\pm$ 0.09  \\
            \midrule       
            Tacotron 2     &       3.58 $\pm$ 0.13 &     3.50 $\pm$ 0.13  \\
            VAE-Tacotron   &       3.76 $\pm$ 0.12 &     3.47 $\pm$ 0.13  \\
            VAE w/ MC      &       3.61 $\pm$ 0.13 &     3.41 $\pm$ 0.13  \\
            VAE w/ CA      &       3.42 $\pm$ 0.13 &     3.30 $\pm$ 0.14  \\
            ReCAB-VAE      &  \bf{3.97 $\pm$ 0.12} & \bf{3.71 $\pm$ 0.13} \\
            \bottomrule  
        \end{tabular}
    \end{table}
    
    In this test, raters were directed to consider the overall quality and naturalness of the speech signals, and assessed them on a 5-point scale with 1-point granularity. Each rater listened to and rated 18 sentences for each system. In order to provide a reference score, we included real human voices and the waveforms which were directly reconstructed from ground truth mel-spectrograms (shown as Recorded and GT-Mel in Table \ref{tab:mos}, respectively). We used Tacotron 2 with and without Gaussian VAE \citep{kingma2013auto} as the baseline. The results are summarized in Table \ref{tab:mos}.
    
    For all of the models except the recorded signal, the total MOS is slightly lower than the corresponding neutral voice MOS.
    Applying a continuous VAE (VAE-Tacotron) results in better scores as a result of the flexibility in controlling continuous latent representation. On the other hand, when the discrete VAE is augmented and trained with the KLD from categorical approximation (CA) or stochastic estimation (MC), the results do not improve over the baseline Tacotron 2 model. 
    When the KLD is approximated by the normal categorical distribution (CA), the naturalness and quality are even worse than the baseline with no latent encoder. This is due to the inaccurate values from the CA method, which makes the training objective unable to manage the trade-off between reconstruction and regularization. 
    In the case of the MC method, performance slightly improves, but is still worse than the continuous-only VAE-Tacotron. This is because of the halved dimension of the continuous latent representation and the poor quality of the discrete latent representation resulting from the inaccurate KLD estimation.
    
    ReCAB shows the best results, which are a significant improvement over the baseline. Similar to the VAE with MC estimation, the training objective with ReCAB is a lower bound of the variational bound. Since ReCAB provides accurate values for the KLD, ReCAB-VAE is able to synthesize natural and high-fidelity speech samples. This improvement is obtained without any architectural changes, showing that ReCAB acts as a better metric for estimating the KLD of the relaxed categorical distribution.
    
    \paragraph{ABX preference test.}    

    \begin{figure}[t]
        \centering
        \includegraphics[width=\textwidth]{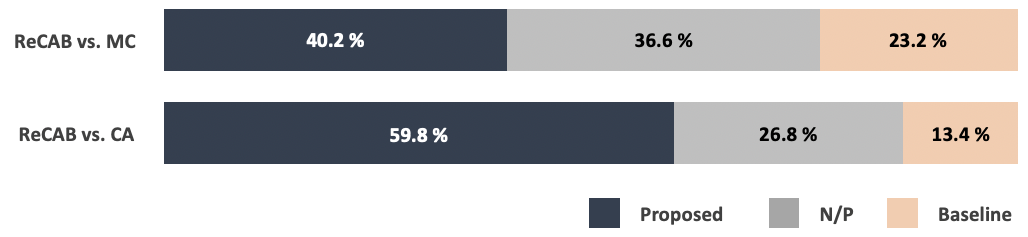}
        \caption{ABX preference test results. \textbf{Proposed} in the figure indicates the model using ReCAB-VAE, and \textbf{N/P} stands for no preference. Samples from the proposed model are preferred over those from both baseline models.}
        \label{fig:prefer}
    \end{figure}
    
    In the ABX preference test, two speech samples of the same sentence and same emotion from different models were presented to raters, who listened to both samples and were asked to make one of the following choices: (A) sample A is preferred, (B) sample B is preferred, or (X) no preference. 
    To evaluate emotional expression and tonal variations, raters considered both emotional expressiveness and naturalness, but were directed to prioritize emotional expression.
    Speech samples from ReCAB-VAE were shown to have higher preference percentages over both the VAE with CA and VAE with MC in these head-to-head comparisons.
    The results of the test are presented in Figure \ref{fig:prefer}.

\section{Conclusion}
\label{sec:conc}
    In this paper, we have proposed a relaxed categorical analytic bound (ReCAB) with a closed-form solution for estimating the Kullback-Leibler divergence (KLD) of a relaxed categorical distribution. Since ReCAB can be solved analytically, it does not depend on a sampling procedure for latent variables and is able to provide a reliable estimation of the KLD. 
    We also proposed a discriminative prior, which enables discriminative learning in variational autoencoders without an additional loss function.
    To evaluate the performance of the proposed metric and prior distribution, we adopted them in a variational autoencoder framework, which we refer to as ReCAB-VAE.
    In an emotional speech synthesis task, ReCAB-VAE was able to generate high fidelity speech samples with rich emotional expressions, and outperformed baseline models in terms of both speech quality and emotion despite only a change to the training objective.
    This indicates that ReCAB is an effective metric for training models that embed latent representations in both continuous and discrete latent spaces, and can improve their performance in various other tasks.
    
\newpage

\bibliographystyle{unsrtnat}  
\bibliography{reference_file}
\medskip
\small

\renewcommand{\thesection}{\Alph{section}}
\renewcommand{\thetable}{s\arabic{table}}
\renewcommand{\thefigure}{s\arabic{figure}}
\renewcommand{\theequation}{s.\arabic{equation}}

\setcounter{section}{0}
\setcounter{table}{0}
\setcounter{figure}{0}
\setcounter{equation}{0}

\newpage
\section*{Supplementary Material}

    \section{Detailed Derivation of ReCAB}
    \subsection{Detailed Derivation} \label{Ax:KLD-Derivation}
        We start from the definition of the KLD of relaxed categorical distribution.
        \begin{align} \begin{split}
            \KLDGS \big( \q(z|x) \| \p(z) \big) 
                &= \EE_z \big[ -(n-1)\log\llamb + \sumN{k}\big( (\log\alpha_k - \lambda\log z_k) - (\log a_k - l \log z_k) \big) \\
                & \qquad -n\lse{i}{\log \alpha_i - \lambda \log z_i} +n\lse{i}{\log a_i - l \log z_i} \big],
        \end{split} \end{align}
        where $\alpha$ and $\lambda$ are parameters of posterior distribution $\q(z|x)$, and $a$ and $l$ are for prior $\p(z)$.
        From the definition of the $k$-th element of relaxed categorical random variable $z_k$, we have:
        \begin{align} \begin{split}
            \log \alpha_k - \lambda \log z_k &= - g_k + \lambda \lse{i}{\z{i}} \\
            \log a_k - l \log z_k &= \log a_k - \llamb \log \alpha_k - \llamb g_k + l \lse{i}{\z{i}},
        \end{split} \end{align}
        where $g_k$ in here is i.i.d. standard Gumbel noise. 
        Since the $\lse{i}{\cdot}$ term is constant with respect to $k$, following equations can be derived:
        \begin{align} \begin{split}
            \lse{k}{\log \alpha_k - \lambda \log z_k} &= \lse{k}{-g_k} + \lambda B \\
            \lse{k}{\log a_k - l \log z_k} &= \lse{k}{A_k - \llamb g_k} + l B,
        \end{split} \end{align}
        where $A_k=\log a_k - \llamb\log\alpha_k$ and $B=\lse{i}{\z{i}}$.
        Applying the above equations, the KLD have the following form:
        \begin{align} \begin{split}
            \KLDGS \big( \q(z|x) \| \p(z) \big)
                &= \EE_g \big[ -(n-1)\log\llamb + \sumN{k}\big\{ \big( -g_k + \lambda B \big) - \big( A_k - \llamb g_k + lB \big) \big\} \\
                    & \qquad {} - n \cdot \big( \lse{k}{-g_k} + \lambda B - \lse{k}{A_k - \llamb g_k} - l B \big) \big]. \\
                &= -(n-1)\log\llamb - \sumN{k} A_k - (1-\llamb) \EE_g\big[\sumN{k}g_k\big] \\
                    & \qquad {} - n\EE_g\big[\lse{k}{-g_k}\big] + n\EE_g\big[ \lse{k}{A_k - \llamb g_k} \big].
        \end{split} \end{align}
        From Jensen's inequality, we can derive the bound of the last term in above equation.
        \begin{align} \begin{split}
            \EE_g \big[ \lse{k}{A_k - \llamb g_k} \big]
                &\le \log \big( \EE_g \big[ \sumN{k} \exp ( A_k - \llamb g_k ) \big] \big) \\
                &= \lse{k}{A_k} + \log \EE_g \big[ \exp (-\llamb g_k) \big] \big)
        \end{split} \end{align}
        Now, we have the inequality for the KLD as following:
        \begin{align} \begin{split}
            \KLDGS \big( \q(z|x) \| \p(z) \big)
                &\le -(n-1) \log \llamb - \sumN{k} \big( A_k - \lse{k}{A_k} \big) - (1-\llamb) \EE_g \big[ \sumN{k} g_k \big] \\
                & \qquad {} - n \EE_g \big[ \lse{k}{-g_k}\big] + n \log \EE_g \big[ \exp (-\llamb g_k) \big].
        \end{split} \end{align}
        The expectation terms in above equation have closed-form solutions.
        
        \paragraph{First expectation: $\EE_g \big[ \sumN{k} g_k \big]$.}
        Since the expectation of sum of random variables is equivalent to sum of expectations,
        \begin{align} \label{eq:first}
            \EE_g \big[ \sumN{k} g_k \big] = n \EE_g \big[ g \big] = n \gamma,
        \end{align}
        where $\gamma \approx 0.57721$ is Euler-Mascheroni constant.
        
        \paragraph{Second expectation: $\EE_g\big[ \lse{k}{-g_k}\big]$.}
        To evaluate this expectation, we first prove that $h=\exp (-g)$ follows exponential distribution. From the CDF of Gumbel distribution $F_G (g') = \exp (\exp(-g'))$, the CDF of $h$ is
        \begin{align} \begin{split}
            F_H (h') &= P(h \le h') = P(\exp(-g) \le h') = P(g \ge -\log h') \\
                &= 1 - P(g \le -\log h') = 1 - F_G(-\log h') \\
                &= 1 - \exp(-h'),
        \end{split} \end{align}
        which is a CDF of exponential distribution with scale parameter of 1, $\text{Exp}(1)$. The summation over $h$, i.e., $\sumN{k}h_k = \sumN{k}\exp(-g_k)$, follows gamma distribution, whose logarithmic expectation is
        \begin{align} \begin{split} \label{eq:second}
            \EE_g \big[ \lse{k}{-g_k} \big] = \psi (n),
        \end{split} \end{align}
        where $\psi (n)$ is a digamma function. The digamma function is a log-derivative of gamma function $\Gamma(\cdot)$ and is calculated as $-\gamma + \sum_{k=1}^{n-1} \frac{1}{k}$ for $n \in \mathbb{N}$. 
        
        \paragraph{Third expectation: $\EE_g \big[ \exp (-\llamb g_k) \big]$.}
        The exponential of weighted Gumbel random variable, $w = \exp (-\llamb g) = h^{\llamb}$, follows Weibull distribution. The CDF of $w$ is
        \begin{align} \begin{split}
            F_W (w') &= P (w \le w') = P(h^{\llamb} \le w') = P(h \le (w')^{\frac{\lambda}{l}}) \\
                &= 1 - \exp (-(w')^{\frac{\lambda}{l}}),
        \end{split} \end{align}
        which is a CDF of $\text{Weibull}(1, \frac{\lambda}{1})$. The expectation of the distribution is
        \begin{align} \label{eq:third}
            \EE_g \big[ \exp (-\llamb g_k) \big] = \Gamma(1+\llamb).
        \end{align}
        
        \paragraph{Resulting ReCAB.} From the results in Eq. (\ref{eq:first}, \ref{eq:second}, and \ref{eq:third}), the resulting bound of the KLD is
        \begin{align} \begin{split} \label{eq:ub}
            \KLDGS \big( \q(z|x) \| \p(z) \big)
                &\le -(n-1) \log \llamb + n \big( \gamma \llamb - \sum_{k=1}^{n-1}\frac{1}{k} + \log \Gamma(1+\llamb) \big) \\
                & \qquad {} - \sumN{k}\big( A_k - \lse{k}{A_k} \big). \\
                &= D_{ReCAB} \big( \q(z|x) \| p(z) \big)
        \end{split} \end{align}
    
    \subsection{Tightness of ReCAB}
        Introducing auxiliary mass $\sumN{k}p_k=1$, we have following lower bound.
        \begin{align} \begin{split}
            \EE_g \big[ \lse{k}{A_k - \llamb g_k} &= \EE_g \big[ \log \sumN{k} p_k \frac{\exp (A_k - \llamb g_k)}{p_k} \big] \\
                &\ge \sumN{k}p_k \EE_g \big[ \log \frac{\exp (A_k - \llamb g_k)}{p_k} \\
                &= \sumN{k}p_k \big( A_k -\gamma\llamb -\log p_k).
        \end{split} \end{align}
        Maximizing this bound with respect to $p_k$ results in the selection of $p_k=\text{softmax}(A_k-\gamma\llamb)$. Applying this value, the bound is
        \begin{align} \begin{split}
            \EE_g \big[ \lse{k}{A_k - \llamb g_k} &\ge \lse{i}{A_i -\gamma\llamb}.
        \end{split} \end{align}
        Note that this bound relates to the lower bound of the KLD. The difference between the upper bound in Eq. (\ref{eq:ub}) and this bound is
        \begin{align} \begin{split}
            n\cdot\big( \log\Gamma(1+\llamb) - \gamma\llamb \big), 
        \end{split} \end{align}
        which is a constant with respect to logit parameter. This is the interval between upper and lower bound of desired KLD, or the maximal gap between proposed ReCAB and true KLD, equivalently.

    \vspace{2in}
    \section{Additional Experimental Results}
        
    \subsection{Sample spectrograms}
        \begin{figure}[h]
            \centering
            \includegraphics[width=0.8\textwidth]{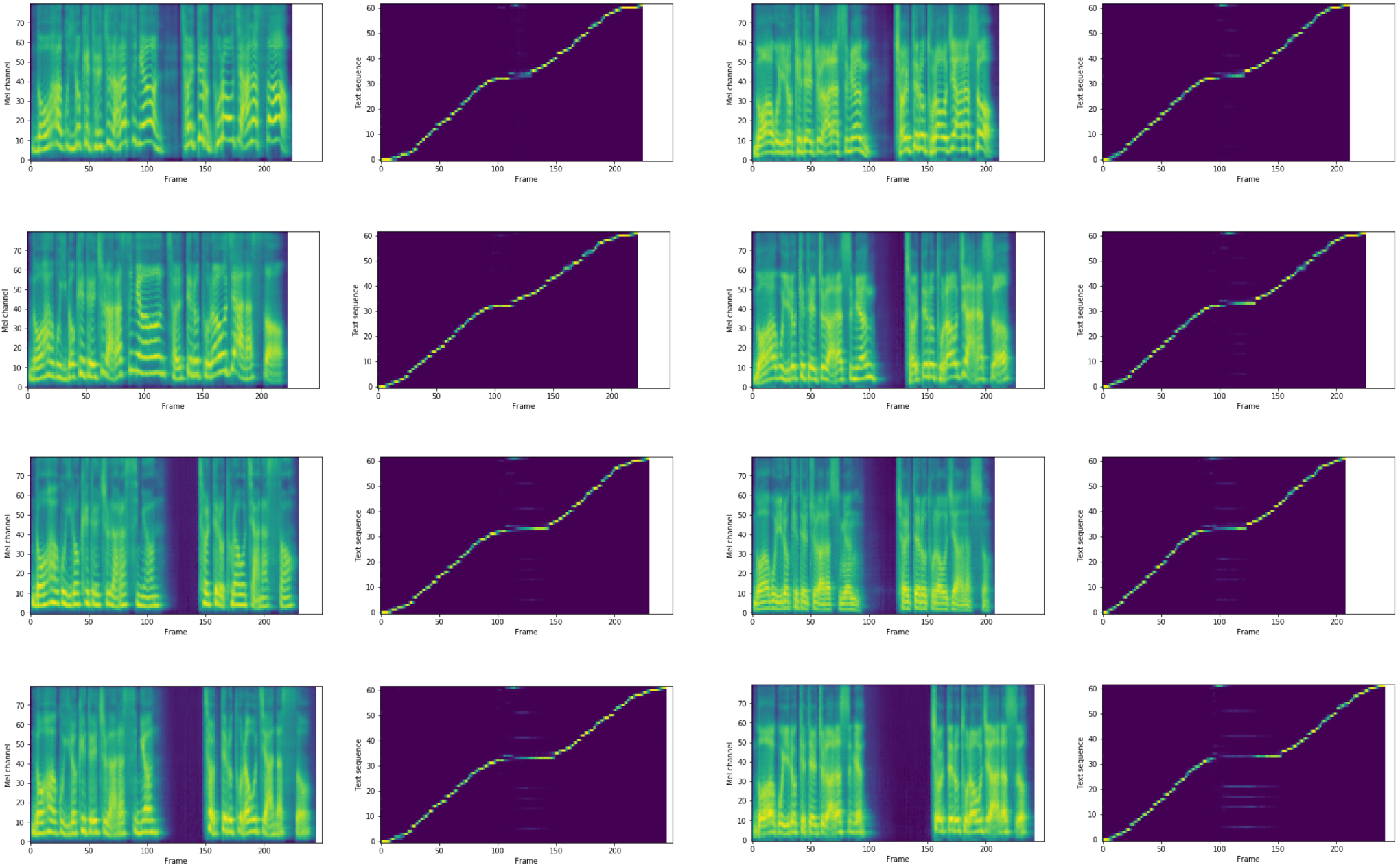}
            \caption{Sample spectrogram and alignment plots of (\textit{left}) female speaker and (\textit{right}) male speaker. Each row is for angry, happy, neutral and sad voices, respectively.}
        \end{figure}
    
    \newpage
    \subsection{Heatmaps of KLD estimations}
        \begin{figure}[h]
            \centering
            \includegraphics[width=0.8\textwidth]{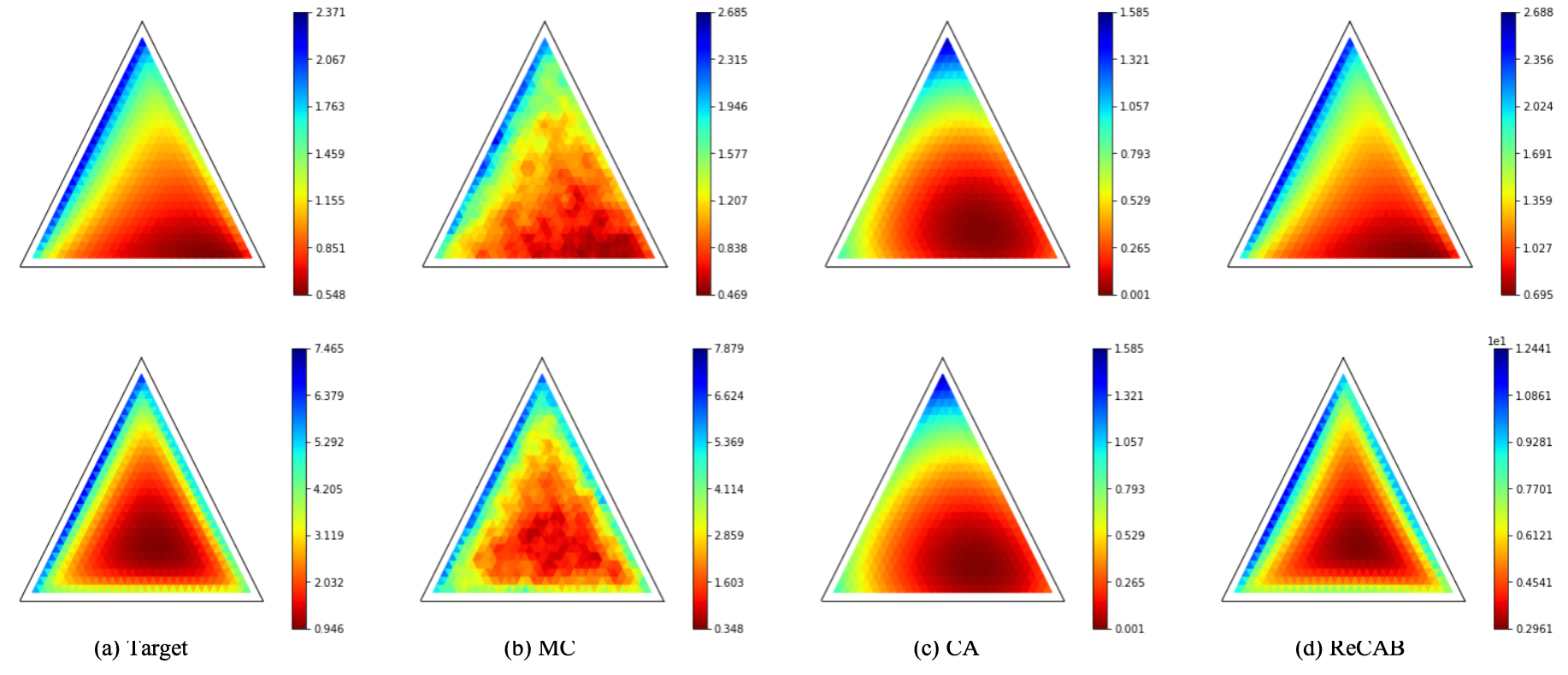}
            \caption{Additional results related to Figure \ref{fig:heat-div} in the main paper. In each row, temperature parameters of target distributions are set to (\textit{top}) 0.5 and (\textit{bottom}) 2, while proposal distribution temperatures are all set to 1. Note that CA does not use the temperature parameter.}
        \end{figure}
        
    \subsection{Density plots}
        \begin{figure}[h]
            \centering
            \includegraphics[width=0.8\textwidth]{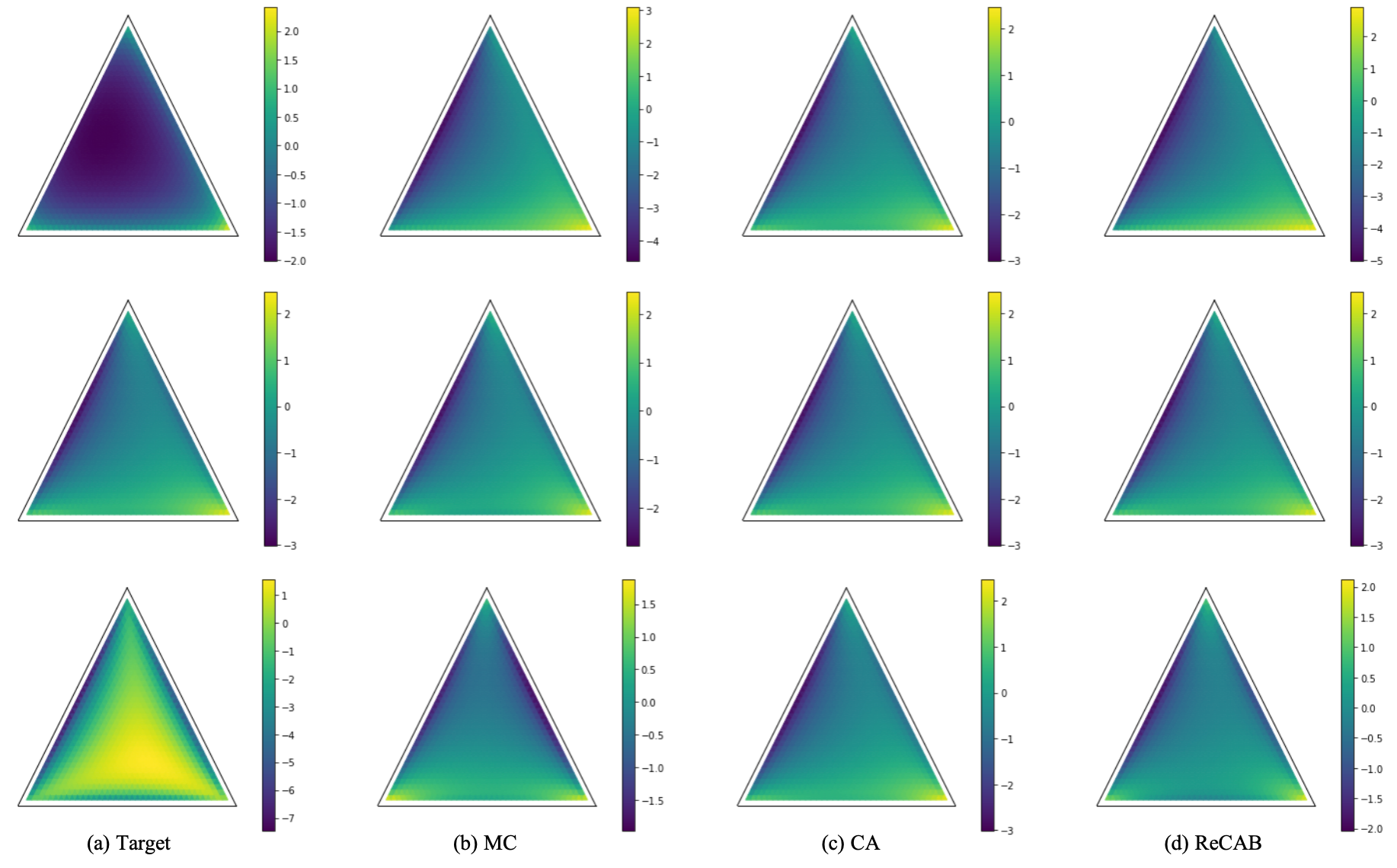}
            \caption{Additional results related to Figure \ref{fig:heat-density} in the main paper. In each row, temperature parameters of target distributions are set to (\textit{top}) 0.5, (\textit{middle}) 1, and (\textit{bottom}) 2, while temperatures of all others equal to 1. Note that same densities are shown in the second row.}
        \end{figure}
    
    \newpage
    \section{Architectures and Settings for Emotional Speech Synthesis}
\begin{figure}[h]
            \centering
            \includegraphics[width=0.8\textwidth]{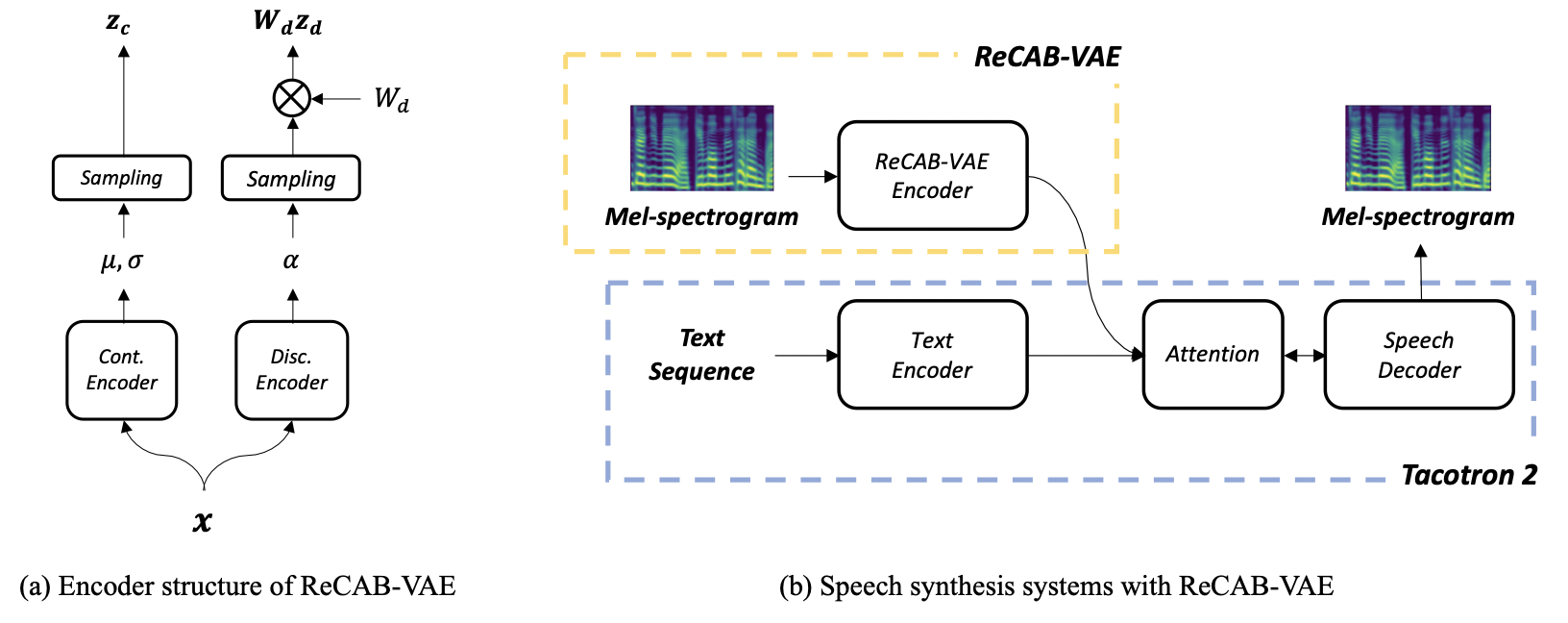}
            \caption{Structure of (a) ReCAB-VAE and (b) ReCAB-based emotional TTS system.}
            \label{fig:arch}
        \end{figure}

        Table \ref{tab:arch} summarizes architectural details, and Figure \ref{fig:arch} illustrates the block diagram of the proposed system.
        The continuous and discrete embeddings are first obtained by the ReCAB-VAE based encoder, after which the embeddings are added to the output embedding of the text encoder in the Tacotron2 framework.
        \begin{table}[h] 
            \caption{Architectural details of the emotional TTS system based on ReCAB-VAE.}
            \vspace{4pt}
            \label{tab:arch}
            \centering
            \begin{tabular}{l l l}
                \toprule        
                Model       & Module            & Hyperparameters \\
                \midrule        
                Data        & Input             & Text embedding (512) \\
                            & Output            & Mel-spectrogram \\
                            &                   & (80, 64 ms frame size with 16 ms hop) \\
                Tacotron 2  & Encoder           & 3 Conv. layers (512) \\
                            &                   & $\rightarrow$ Bi-LSTM (512) \\
                            & Attention type    & Location sensitive attention \\
                            & Decoder           & 2-layer Pre-Net (FC, 256-dim) \\
                            &                   & $\rightarrow$ AttentionRNN (1024, LSTM) \\
                            &                   & $\rightarrow$ DecoderRNN (512, LSTM) \\
                            &                   & $\rightarrow$ FC layer (80) \\
                            &                   & $\rightarrow$ Post-Net (80-512*3-80, 5 Conv.) \\
                ReCAB-VAE   & Cont.Encoder      & 6 Conv. layers (16-16-32-32-64-64 channels) \\
                            &                   & $\rightarrow$ GRU (256) \\
                            &                   & $\rightarrow$ FC (256) \\
                            & Disc.Encoder      & 6 Conv. layers (16-16-32-32-64-64 channels) \\
                            &                   & $\rightarrow$ GRU (256) \\
                            &                   & $\rightarrow$ FC layer (256) \\
                \bottomrule  
            \end{tabular}
        \end{table}

\end{document}